\documentstyle[twoside,fleqn,espcrc2,epsf]{article}

\newcommand{\ainv}{{a^{-1}}}
\newcommand{\be}{\begin{equation}}
\newcommand{\ee}{\end{equation}}

\newcommand{\alphaMSMZ}{\alpha_{\overline{MS}}[M_{Z^0}]}
\newcommand{\alphaMS}{\alpha_{\overline{MS}}}

\newcommand{\AmS}{{\protect\the\textfont2
  A\kern-.1667em\lower.5ex\hbox{M}\kern-.125emS}}

\title{ Quarkonium Physics and $\alpha_{strong}$ from  Quarkonia }

\author{
        Junko Shigemitsu 
        \address{Physics Department, 
        The Ohio State University, 
        Columbus, Ohio 43210, USA.}
        \thanks{Plenary Talk presented at LATTICE'96.}
                           }
       
\begin{document}

\begin{abstract}

Recent results in Quarkonia are reviewed, including updates on spectroscopy 
and $\alpha_s$, and a first look at quarkonium annihilation decays.

\end{abstract}

\maketitle

\section{Introduction}
Quarkonia serve as good testing ground 
for QCD and for theoretical approaches to 
QCD, including the lattice, 
due to their rich spectroscopy and decay phenomenology. 
Furthermore,  heavy-heavy and heavy-light systems,
 are playing a crucial role in probing the electroweak 
sector of the Standard Model.  Hence, it is very important that we have 
reliable well tested methods for simulating heavy quarks.

Both four component and two component nonrelativistic (NRQCD) fermions are 
being used to simulate heavy quarks \cite{fnal,nrqcd}. 
 Most of the theory has been reviewed 
 in previous lattice meetings.  Only a few comments on recent developments 
in the four component approach will be made here.  
The first formulation of four component heavy quarks on the lattice was
 developed by the Fermilab group \cite{fnal}. 
 To date this approach  has been 
implemented at relatively low order. 
 At this level one works with the 
same set of parameters as in the clover action for light quark simulations, 
namely the Wilson hopping parameter $\kappa$ and the clover coefficient $c$.
For heavy quarks, truncation to these two parameters, does not allow the 
coefficient of the $p^4$ term in a low momentum dispersion expansion,
$ E = M_1 + \vec{p}^2/2M_2 + O(p^4) $ to come out correctly. 
What implications this has for 
$M_1^{hadron}$ versus $M_2^{hadron}$ of composite particles,
is discussed by Andreas Kronfeld in a poster 
session at this conference \cite{kronfeld}. 
For hadrons containing heavy quarks, it is $M_2^{hadron}$ that should 
be identified with the physical mass. 
In order to minimize uncertainties associated with 
$M_2^{hadron}$ and in fixing $\kappa$ it is 
important that the $p^4$ contribution 
in the action be improved.  Problems with $M_2$ in heavy-heavy and 
heavy-light hadrons using a straight tadpole improved clover action,
 were pointed out a year ago at Melbourne by Sara Collins and 
John Sloan \cite{m1m2}.

Other four component approaches to heavy quarks have now appeared. 
 Alford-Klassen-Lepage have a D234(2/3) action that goes beyond 
clover and removes $O(a^2)$ errors at tree level. Working with 
anisotropic lattices, they find good relativistic dispersion relation 
and spectroscopy results 
even for heavy quarks around charm \cite{akl}. 
 The MIT-BU group has implemented 
a ``perfect action'' for fermions truncated to couplings within a 
hypercube and also find encouraging results for charm \cite{mitbu}.

 Due to time constraints, for the rest of the talk I will concentrate on 
quarkonium physics results rather than formalism and refer the reader to 
parallel and poster sessions for more theory.

\section{Brief Overview of Spectroscopy}

  The first test of  any lattice method for heavy fermions, is to see how 
well the observed quarkonium spectrum can be reproduced.  Figures 1 \& 2 
summarize Upsilon spectroscopy results by several groups using both 
quenched ($n_f = 0$) and dynamical ($n_f = 2$) configurations.  Tadpole 
improved clover (Fermilab \& SCRI) and nonrelativistic fermions 
with improvements through $O(Mv^4)$ (NRQCD) have been employed. 
 Two parameters were fixed by experiment, the bare quark mass and 
the lattice spacing.  Fermilab and SCRI use the S-P splitting to fix 
$\ainv$.  The NRQCD collaboration determines $\ainv$ from both the S-P and 
the 1S-2S splittings and uses an average in the plots.  A first glance 
at Fig. 1 suggests good agreement between simulations and experiment.  
A closer look reveals, for instance, that for the NRQCD data neither the 
1P nor the 2S state lies on the experimental line.  This is 
particularly noticeable for the $ n_f = 0$ results and is a reflection of a
quenching effect that leads to different quantities giving different 
$\ainv$'s for the wrong number of dynamical flavors if these quantities 
 are governed by different characteristic energy scales.  
We will see later, that such details are important in precise 
determinations of $\alpha_s$.

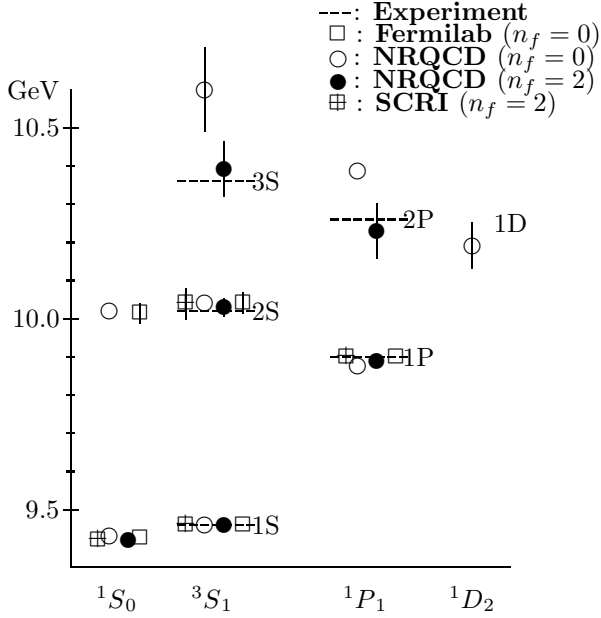
\begin{figure}[t]
\begin{center}
\setlength{\unitlength}{.02in}
\begin{picture}(130,140)(10,930)
\put(15,935){\line(0,1){125}}
\multiput(13,950)(0,50){3}{\line(1,0){4}}
\multiput(14,950)(0,10){10}{\line(1,0){2}}
\put(12,950){\makebox(0,0)[r]{9.5}}
\put(12,1000){\makebox(0,0)[r]{10.0}}
\put(12,1050){\makebox(0,0)[r]{10.5}}
\put(12,1060){\makebox(0,0)[r]{GeV}}
\put(15,935){\line(1,0){115}}


\multiput(80,1080)(3,0){3}{\line(1,0){2}}
\put(89,1080){\makebox(0,0)[l]{: {\bf Experiment}}}
\put(81,1074){\makebox(0,0)[l]{$\,\Box $ : {\bf Fermilab 
$(n_f = 0)$}}}
\put(85,1068){\makebox(0,0)[tl]{\circle{4}}}
\put(89,1068){\makebox(0,0)[l]{: {\bf NRQCD $(n_f = 0)$}}}
\put(85,1062){\makebox(0,0)[tl]{\circle*{4}}}
\put(89,1062){\makebox(0,0)[l]{: {\bf NRQCD $(n_f = 2)$}}}
\put(81,1056){\makebox(0,0)[l]{$\,\Box $ : {\bf SCRI 
$(n_f = 2)$}}}
\put(81,1056.7){\makebox(0,0)[l]{$\,+ \;\;$  }}

\put(27,930){\makebox(0,0)[t]{${^1S}_0$}}
\put(25,943.1){\circle{4}}
\put(30,942){\circle*{4}}
\put(33,942){\makebox(0,0){$\Box$}}
\put(22,941.5){\makebox(0,0){$\Box$}}
\put(22,942.5){\makebox(0,0){$+$}}

\put(25,1002){\circle{4}}
\put(33,1001){\makebox(0,0){$\Box$}}
\put(33,1001.4){\line(0,1){2.6}}
\put(33,1001.4){\line(0,-1){2.6}}

\put(52,930){\makebox(0,0)[t]{${^3S}_1$}}
\put(66,946){\makebox(0,0){1S}}
\multiput(43,946)(3,0){7}{\line(1,0){2}}
\put(50,946){\circle{4}}
\put(55,946){\circle*{4}}
\put(60,945.5){\makebox(0,0){$\Box$}}
\put(45,945.5){\makebox(0,0){$\Box$}}
\put(45,946.5){\makebox(0,0){$+$}}

\put(66,1002){\makebox(0,0){2S}}
\multiput(43,1002)(3,0){7}{\line(1,0){2}}
\put(50,1004.1){\circle{4}}
\put(55,1003){\circle*{4}}
\put(55,1004){\line(0,1){1.4}}
\put(55,1002){\line(0,-1){1.4}}
\put(60,1003.5){\makebox(0,0){$\Box$}}
\put(60,1004){\line(0,1){2.7}}
\put(60,1004){\line(0,-1){2.7}}
\put(45,1003.5){\makebox(0,0){$\Box$}}
\put(45,1004.3){\makebox(0,0){$+$}}
\put(45,1003.9){\line(0,1){4}}
\put(45,1003.9){\line(0,-1){4}}

\put(66,1036){\makebox(0,0){3S}}
\multiput(43,1036)(3,0){7}{\line(1,0){2}}
\put(50,1060){\circle{4}}
\put(50,1060){\line(0,1){11}}
\put(50,1060){\line(0,-1){11}}
\put(55,1039.1){\circle*{4}}
\put(55,1039.1){\line(0,1){7.2}}
\put(55,1039.1){\line(0,-1){7.2}}

\put(92,930){\makebox(0,0)[t]{${^1P}_1$}}

\put(106,990){\makebox(0,0){1P}}
\multiput(83,990)(3,0){7}{\line(1,0){2}}
\put(90,987.6){\circle{4}}
\put(95,989){\circle*{4}}
\put(100,989.5){\makebox(0,0){$\Box$}}
\put(87,989.5){\makebox(0,0){$\Box$}}
\put(87,990.3){\makebox(0,0){$+$}}

\put(106,1026){\makebox(0,0){2P}}
\multiput(83,1026)(3,0){7}{\line(1,0){2}}
\put(90,1038.7){\circle{4}}
\put(95,1023){\circle*{4}}
\put(95,1023){\line(0,1){7.2}}
\put(95,1023){\line(0,-1){7.2}}

\put(130,1025){\makebox(0,0){1D}}
\put(120,930){\makebox(0,0)[t]{${^1D}_2$}}
\put(120,1019.2){\circle{4}}
\put(120,1019.2){\line(0,1){6}}
\put(120,1019.2){\line(0,-1){6}}

\end{picture}
\end{center}

\caption{{\bf  $\Upsilon$ Spectrum }
}
\end{figure}

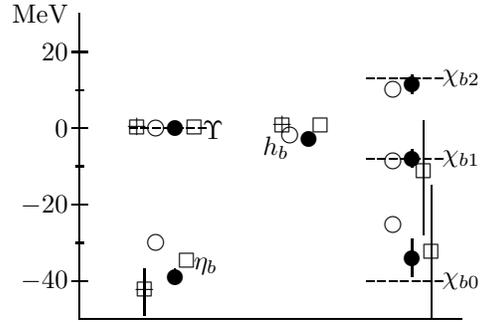
\begin{figure}
\begin{center}
\setlength{\unitlength}{.02in}
\begin{picture}(100,80)(15,-50)

\put(15,-50){\line(0,1){80}}
\multiput(13,-40)(0,20){4}{\line(1,0){4}}
\multiput(14,-40)(0,10){7}{\line(1,0){2}}
\put(12,-40){\makebox(0,0)[r]{$-40$}}
\put(12,-20){\makebox(0,0)[r]{$-20$}}
\put(12,0){\makebox(0,0)[r]{$0$}}
\put(12,20){\makebox(0,0)[r]{$20$}}
\put(12,30){\makebox(0,0)[r]{MeV}}
\put(15,-50){\line(1,0){100}}


\multiput(28,0)(3,0){7}{\line(1,0){2}}
\put(50,2){\makebox(0,0)[t]{$\Upsilon$}}
\put(35,0){\circle{4}}
\put(40,0){\circle*{4}}
\put(45,-0.5){\makebox(0,0){$\Box$}}
\put(30,-0.5){\makebox(0,0){$\Box$}}
\put(30,0.5){\makebox(0,0){$+$}}

\put(48,-34){\makebox(0,0)[t]{$\eta_b$}}
\put(35,-29.9){\circle{4}}
\put(40,-39){\circle*{4}}
\put(40,-39){\line(0,1){2}}
\put(40,-39){\line(0,-1){2}}
\put(43,-35.3){\makebox(0,0){$\Box$}}
\put(32,-43){\makebox(0,0){$\Box$}}
\put(32,-42.3){\makebox(0,0){$+$}}
\put(32,-43){\line(0,1){6}}
\put(32,-43){\line(0,-1){6}}

\put(63,-5){\makebox(0,0)[l]{$h_b$}}
\put(70,-1.8){\circle{4}}
\put(75,-2.9){\circle*{4}}
\put(75,-2.9){\line(0,1){1.2}}
\put(75,-2.9){\line(0,-1){1.2}}
\put(78, 0.){\makebox(0,0){$\Box$}}
\put(68, 0.){\makebox(0,0){$\Box$}}
\put(68, 0.9){\makebox(0,0){$+$}}

\multiput(90,-40)(3,0){7}{\line(1,0){2}}
\put(110,-40){\makebox(0,0)[l]{$\chi_{b0}$}}
\put(97,-25.1){\circle{4}}
\put(102,-34){\circle*{4}}
\put(102,-34){\line(0,1){5}}
\put(102,-34){\line(0,-1){5}}
\put(107,-33){\makebox(0,0){$\Box$}}
\put(107,-33){\line(0,1){18}}
\put(107,-33){\line(0,-1){17}}

\multiput(90,-8)(3,0){7}{\line(1,0){2}}
\put(110,-8){\makebox(0,0)[l]{$\chi_{b1}$}}
\put(97,-8.6){\circle{4}}
\put(102,-7.9){\circle*{4}}
\put(102,-7.9){\line(0,1){2.4}}
\put(102,-7.9){\line(0,-1){2.4}}
\put(105,-12){\makebox(0,0){$\Box$}}
\put(105,-12){\line(0,1){14}}
\put(105,-12){\line(0,-1){16}}

\multiput(90,13)(3,0){7}{\line(1,0){2}}
\put(110,13){\makebox(0,0)[l]{$\chi_{b2}$}}
\put(97,10.2){\circle{4}}
\put(102,11.5){\circle*{4}}
\put(102,11.5){\line(0,1){2.4}}
\put(102,11.5){\line(0,-1){2.4}}

\end{picture}
\end{center}
 \caption{{\bf $\Upsilon$ Spin Splittings }: Symbols have the 
same meaning as in Figure 1.}
\end{figure}

Quenching effects are also observable in the spin splittings of Fig.2 .  They 
tend to underestimate spin splittings.  Hyperfine and fine-struture 
splittings are discussed in more detail in later sections.  
Fig. 3  shows lattice spectrum results for Charmonium.

\begin{figure}
\begin{center}
\setlength{\unitlength}{.02in}
\begin{picture}(200,140)(2,280)

\put(15,285){\line(0,1){120}}
\multiput(13,300)(0,50){3}{\line(1,0){4}}
\multiput(14,290)(0,10){10}{\line(1,0){2}}
\put(12,300){\makebox(0,0)[r]{3.0}}
\put(12,350){\makebox(0,0)[r]{3.5}}
\put(12,400){\makebox(0,0)[r]{4.0}}
\put(12,410){\makebox(0,0)[r]{GeV}}
\put(15,285){\line(1,0){130}}


     \put(30,280){\makebox(0,0)[t]{$\eta_c$}}
\multiput(23,298)(3,0){5}{\line(1,0){2}}
     \put(29,300){\circle{4}}
     \put(33,301){\makebox(0,0){$\Box$}}
     \put(26,300.3){\makebox(0,0){$\Box$}}
     \put(26,301.1){\makebox(0,0){$+$}}

     \put(28,368){\circle{4}}
     \put(28,368){\line(0,1){6}}
     \put(28,368){\line(0,-1){6}}
     \put(32,363.5){\makebox(0,0){$\Box$}}
     \put(32,363.5){\line(0,1){4.4}}
     \put(32,363.5){\line(0,-1){3.5}}

     \put(50,280){\makebox(0,0)[t]{$J/\Psi$}}
\multiput(43,310)(3,0){5}{\line(1,0){2}}
     \put(49,309){\circle{4}}
     \put(53,308.5){\makebox(0,0){$\Box$}}
     \put(46,308.5){\makebox(0,0){$\Box$}}
     \put(46,309.3){\makebox(0,0){$+$}}

\multiput(43,369)(3,0){5}{\line(1,0){2}}
     \put(48,371.5){\circle{4}}
     \put(48,371.5){\line(0,1){8}}
     \put(48,371.5){\line(0,-1){8}}
     \put(52,370){\makebox(0,0){$\Box$}}
     \put(52,370){\line(0,1){4.4}}
     \put(52,370){\line(0,-1){3.1}}
     \put(45,364){\makebox(0,0){$\Box$}}
     \put(45,364.8){\makebox(0,0){$+$}}
     \put(45,364){\line(0,1){5}}
     \put(45,364){\line(0,-1){5}}

     \put(70,280){\makebox(0,0)[t]{$h_c$}}
\multiput(63,352.6)(3,0){5}{\line(1,0){2}}
     \put(69,353){\circle{4}}
     \put(73,352.5){\makebox(0,0){$\Box$}}
     \put(66,352.5){\makebox(0,0){$\Box$}}
     \put(66,353.3){\makebox(0,0){$+$}}

     \put(90,280){\makebox(0,0)[t]{$\chi_{c0}$}}
\multiput(83,341.5)(3,0){5}{\line(1,0){2}}
     \put(88,346.4){\circle{4}}
     \put(92,346.2){\makebox(0,0){$\Box$}}
     \put(92,346.7){\line(0,1){1.8}}
     \put(92,346.7){\line(0,-1){2.1}}

     \put(105,280){\makebox(0,0)[t]{$\chi_{c1}$}}
\multiput(98,351)(3,0){5}{\line(1,0){2}}
     \put(103,351.8){\circle{4}}
     \put(107,350.5){\makebox(0,0){$\Box$}}
     \put(107,351) {\line(0,1){2}}
     \put(107,351) {\line(0,-1){1.9}}

     \put(120,280){\makebox(0,0)[t]{$\chi_{c2}$}}
\multiput(113,355.6)(3,0){5}{\line(1,0){2}}
     \put(120,357.2){\circle{4}}

\end{picture}
\end{center}
 \caption{{\bf Charmonium Spectrum }: Symbols have the 
same meaning as in Figure 1.}
\end{figure}
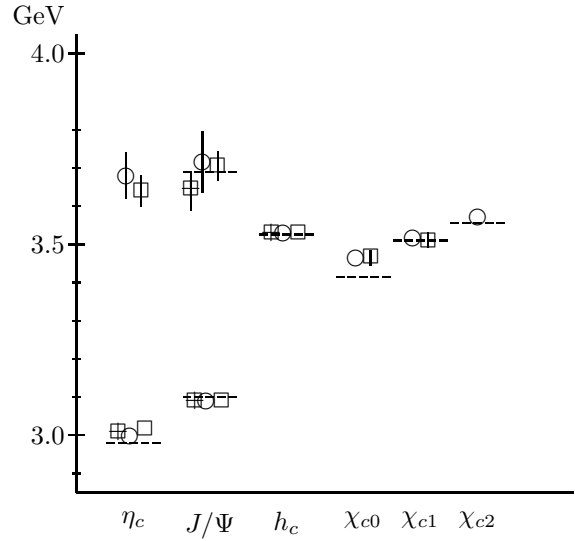

\section{$\alpha_s$ from Quarkonia}

One major goal of lattice gauge theory is to contribute towards 
determination of fundamental parameters of the Standard Model.  Quarkonium 
studies have led to accurate determinations of heavy quark masses 
\cite{mb} and 
of the strong coupling constant \cite{fnalalpha,nrqcdalpha}.   
 $\alpha_s$ estimates are continually being refined and improved 
\cite{weisz}.  We present here a status report.

\subsection {Procedure}

There are three steps involved in determining $ \alpha_s$ from the 
lattice and 
comparing with other determinations. 
(i)  Define a suitable $\alpha_s$ and measure its value.
(ii)  Set the scale at which $\alpha_s$ takes on this particular 
value.
(iii)  Convert to $\alphaMS$.

For step (i), two popular choices are $\alpha_P$ ( P: plaquette )
\cite{lepmac} and 
$\alpha_{SF}$ ( SF: Schroedinger Functional )\cite{sf}.
In quarkonium studies to date one has used $\alpha_P$.  It is defined 
through the $1 \times 1$ Wilson loop as,

$$
-lnW_{1,1} = {4\pi \over 3}\alpha_P({3.4 \over a}) [1 - (1.19
+ 0.07n_f)\alpha_P] $$

\noindent
$\alpha_P$ agrees through order $\alpha^2$ with the more familiar 
$\alpha_V$ of Brodsky-Lepage-Mackenzie.  It is defined so that the above 
relation 
terminates at $\alpha^2$ and 
there are  no higher order corrections.   
Hence the only errors in  measuring the value of  $\alpha_P$ 
 come from statistical errors in $W_{1,1}$ which 
are extremely small. In principle one must ascertain that nonperturbative, 
i.e. non power series, contributions to $W_{1,1}$ are negligible.  
Several tests confirm that any nonperturbative contamination would 
be smaller than one percent \cite{nrqcdalpha}.

 Most of the work and the source of all the uncertainty 
in $\alpha_s$ determinations involve
 steps (ii) and (iii).  Quarkonium studies come into play at step (ii), when 
one sets the scale $ Q = {3.4 \over a}$. by inserting a value for $\ainv$.
Any dimensionful quantity can be used to determine $\ainv$.  
Quarkonium level splittings are particularly suitable since 
systematic errors are easiest to control.

\noindent
\underline{Finite volume : }  Quarkonia, especially $b \overline{b}$, are 
small compared to light hadrons

\noindent
\underline{ $M_Q$ dependence :} There is no need to carry out mass 
extrapolations like the chiral extrapolation in light hadron studies. 
One works directly at or very close to realistic heavy quarks masses. 
Furthermore 
experimentally S-P and 1S-2S splittings in quarkonia are very insensitive 
to $ M_Q$ in the charm to bottom quark region.

\noindent
\underline {Finite a:} Heavy quark actions, such as NRQCD, can and have 
been improved through order $a^2$.  Effects of $a^2$ errors in the Wilson 
glue action have been estimated perturbatively.  Fully improved gluonic 
actions through order $a^2$ and beyond, are starting to be employed 
as well.

\noindent
\underline{Quenching or $ n_f$ dependence :}  This is the most subtle 
systematic error that needs to be understood and corrected for.
Two questions must be settled first; 
1.  What is $n_f^{phys.}$, the relevant number of dynamical quark flavors ?
2. How do quarkonium splittings depend on $m_q^{dyn.}$, the dynamical 
quark mass ?

\noindent
Typical gluon momenta $p_{\Upsilon}$ inside an $\Upsilon$ are between 
0.5 and 1.GeV.  The relation $ m_u,\;  m_d, \; m_s << p_{\Upsilon} < 
m_c$  tells us that $ n_f^{phys.} = 3$ for $\Upsilon$ splitting physics.
This answers the first question.  
It is necessary to ask the second question, because dynamical quarks 
in current simulations 
are heavier than physical up- or down-quarks.
Perturbation theory would indicate that energy level splittings, 
$\Delta E$,  depend quadratically on $m_q^{dyn.}$.  However, due to 
chiral symmetry breaking in QCD, $ \Delta E$ actually 
has a linear $m_q^{dyn}$ dependence \cite{grinroth}.  The relevant 
combination of 
light quark masses is  $( m_u + m_d + m_s) $.  Since all three quark
masses are significantly smaller than $p_{\Upsilon}$, one should be 
able to replace the three unequal mass quarks with three flavors 
of degenerate quarks with $m_{eff} = (m_u + m_d + m_s)/3 \sim m_s/3$.
  $\Upsilon$ physics should not care about  the details of the $m_{u,d} -
 m_s$ mass differences.  
In summary, to undo quenching effects, one needs to extrapolate, 
$ n_f \rightarrow n_f^{phys.} = 3 $  and 
$ m_q^{dyn.} \rightarrow m_s/3 $.

\noindent
The fact that one only needs to extrapolate in 
$m_q^{dyn}$ down to $m_s/3$ and not beyond,  highlights one of the true 
advantages of working with the $\Upsilon$ system.  Unquenching in systems 
where one is more sensitive to $m_{u,d} - m_s$ mass differences, e.g. 
any system where characteristic gluon momenta are of order $\Lambda_{QCD}$ 
or $m_s$ will be more complicated.  Even within quarkonia,  the charmonium 
system should have larger uncertainties associated with 
$m_q^{dyn}$ than $\Upsilon$.

\subsection{ Results}

For the quenched, $n_f = 0$, case several groups have now extracted 
$\ainv$ from quarkonium S-P splittings, which can then be used to set the 
scale in $\alpha_P^{(n_f=0)}[{3.4 \over a}]$.  However, very few $ n_f > 0$ 
results exist to date.  The most accurate numbers come from the NRQCD 
collaboration (using HEMCGC dynamical configurations)
 who have combined this with their $n_f = 0$ numbers and 
carried out the extrapolation to $ n_f^{phys}$ described above
 \cite{nrqcdalpha,nrqcd3}.  Recently 
the SCRI group has calculated an $n_f =2$ $\ainv$ from the $\Upsilon$ 
S-P splitting on the same dynamical configurations, however using 
tadpole improved clover 
 \cite{scri}.  I have taken the liberty to combine their $\ainv$ with 
quenched clover 
results from the Fermilab group to perform the $n_f \rightarrow 
n_f^{phys}$ extrapolation.

Fig. 4 summarizes results for $1/\alpha_P$ versus $ln({3.4 \over a})$.  
This is an updated version of a plot originally made 
 by Matthew Wingate \cite{matthew}.  Only data from the last two 
years are included.  There are several comments to be made.

 The $n_f =0$ charmonium data (open boxes) and $\Upsilon$ data 
(open circles) lie on two separate scaling curves.  This is due to the fact 
that the characteristic energy scale differs between charmonium and 
$\Upsilon$.  In a quenched simulation, $\alpha[Q]$ runs incorrectly 
between the two characteristic scales. When real world data (which 
incorporates correct running of $\alpha_s$) is used to extract 
$\ainv$, a mismatch is found which results in the two scaling curves 
being shifted horizontally with respect to each other.
A similar phenomenon is found by the NRQCD collaboration when they 
compare $\ainv$'s from $\Upsilon$ S-P and 1S-2S splittings.  
It is only after extrapolating to $n_f^{phys.}$ that physics becomes 
independent of which splitting was used ( see below).

The $n_f =2 $ configurations used by NRQCD and SCRI have
$am_q^{dyn.} = 0.01$.  How close is this to the desired $m_s/3$ ?  Based 
on the light staggered spectrum calculations by HEMCGC \cite{hemcgc},
 this $m_q^{dyn.}$ 
lies anywhere between $\sim m_s/2$ (from the kaon) and $\sim m_s$ 
(from the $\phi$). Hence there are large uncertainties in how far 
to extrapolate in $m_q^{dyn.}$.  Furthermore, in order to do the 
extrapolation, data at another value of $m_q^{dyn.}$ are required.  
The NRQCD collaboration has results from $am_q^{dyn.} = 0.025$\cite{nrqcd3}.
  The 
errors in those simulations are still large.  A 1 to 1.5 $\sigma$ difference 
was found between  $am_q^{dyn.}$ = 0.01 and 0.025.  It is not 
clear yet whether true $m_q^{dyn.}$-dependence has been observed.  So, the 
0.025 data has only been used to estimate the error associated with 
$m_q^{dyn.}$.  The central value for $\alpha_p$ is given by the 0.01 result 
and the error is estimated by taking the difference between this and an
extrapolation using the 0.025 data to $am_q^{dyn.} = (0.01)/3$.

\noindent
The NRQCD results after the $n_f \rightarrow 3$ 
extrapolation in $1/\alpha_p$ are,

\[\alpha_p^{(n_f=3)}[8.2GeV] =
\cases{ 
.1948\, (29)(11)(37) \;(S-P) \cr
.1962\, (41)(08)(40) \; (1S-2S)\cr} \]

\noindent
The three errors correspond to statistical, discretization/relativistic 
and to $m_q^{dyn.}$ errors respectively.  
The S-P result is the rightmost $n_f = 3$ (fancy box) data point in 
Fig. 4 ( the 1S - 2S number falls right on top of it).  The 
other four $n_f = 3$ data points correspond from right to left to 
Fermilab/SCRI, CDHW\cite{cdhw} and to two preliminary NRQCD results 
from $\Upsilon$
 and charm 
at lower $\beta$, using UKQCD quenched  and MILC dynamical configurations.

\begin{figure}
\epsfxsize=7.0cm
\epsfbox{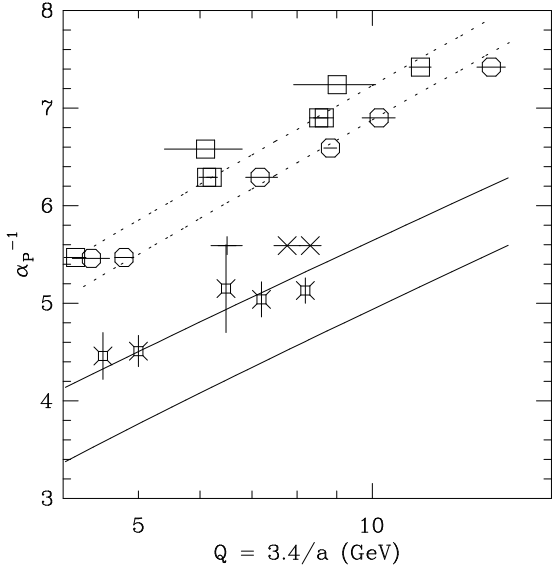}
\caption{
$ 1/\alpha_P$ versus $ ln({ 3.4 \over a})$. 
 $n_f=0$ charm: boxes [18,19,20,22,23]; $n_f=0$ $\Upsilon$: circles 
[18,19,21,14]; 
 $ n_f = 2$ charm: $+$ [23]; $ n_f =2$ $\Upsilon$: $\times$ 
[9,14,15];
 $n_f = 3$ (extrapolated): fancy boxes.
Non lattice determinations of $\alpha_s$
typically find values for $\alphaMSMZ$ 
 between $0.115 \sim 0.125$.  Converting to $\alpha_P^{(n_f=3)}$,
this corresponds to the region between the two full curves in the plot.
The dotted lines are $n_f=0$ three loop scaling curves.
}

\end{figure}

\noindent
Finally, step (iii) involves conversion to the $\overline{MS}$ scheme, 
in order to be able to compare with other determination of $\alpha_s$.  
The conversion formula is,

\begin{eqnarray}
\alpha_{\overline{MS}}^{(n_f)}(Q)& =& \alpha_P^{(n_f)}(e^{5/6}Q)\nonumber\\
&& \times \left[1+\frac{2}{\pi}\alpha_P^{(n_f)}+ C_2(n_f) \;\alpha_P^2
\nonumber   \right]
\end{eqnarray}

\noindent
The two loop coefficient $C_2$ has been calculated by Luescher \& Weisz for 
the quenched theory,  $C_2(n_f=0) = 0.96$
 \cite{lw}.  In the past $C_2 = 0$ was used. 
It now makes more sense to use $C_2 = 0.96$ for central values even for 
$\alpha^{(n_f=3)}$.  The difference between $C_2=0$ and $C_2=0.96$ will 
serve as an estimate of errors due to uncertainties in the $\alpha_p - 
\alpha_{\overline{MS}}$ conversion.

\noindent
New results for $\alphaMSMZ$ exist this year from the NRQCD collaboration, 
who have updated their previous numbers, and from combining 
Fermilab's quenched and SCRI's dynamical $\Upsilon$ data. 
NRQCD quotes (the Luescher-Weisz $C_2$ has been used),

\[\alphaMSMZ =
\cases{ 
.1175\, (11)(13)(19) \;(S-P) \cr
.1180\, (14)(14)(19) \; (1S-2S)\cr} \]

\noindent
The first error now includes both statistical
 and discretization/relativistic 
errors,  the second is due to $m_q^{dyn.}$ and the third comes from 
conversion uncertainties.  The combined Fermilab/SCRI S-P data give,

$$\alphaMSMZ = 0.1159\, (19)(13)(19)  $$

\noindent
This is a very preliminary number based on this author's 
extrapolations. 
Nevertheless it is encouraging that results using two different 
fermion methods are approaching one another.

\noindent
In the future, improvements in $\alpha_s$ from quarkonia should 
come from further studies of $m_q^{dyn.}$ dependence.  Simulations 
with dynamical Wilson configurations or staggered with $n_f > 2$ 
would be welcome.  The largest source of uncertainty for $\alphaMS$ is 
still higher orders in the conversion formula, so $C_2(n_f > 0)$ is 
crucial.  Finally, unquenched lattice $\alpha_s$ determinations 
other than from quarkonia are eagerly awaited.

\section{Quarkonium Annihilation Decays}

Annihilation decays of quarkonia provide another fertile arena for 
lattice investigations.  Work has just begun in this area with first 
results coming from the Argonne group \cite{argonne}. 
 The theory was developed a 
couple of years ago by Bodwin-Braaten-Lepage, using the framework 
of NRQCD\cite{bbl}.
 Annihilation decay rates $\Gamma( Q \overline{Q} \rightarrow 
X )$ ( X : light hadrons, $\gamma \gamma$, $l^+ l^-$),  can be 
expressed as a power series in $1/M_Q$. The coefficients factorize 
into a  short distance perturbative and a non-perturbative part.  
Lattice methods can be applied to obtain the non-perturbative contributions 
that enter as hadronic matrix elements of four fermion operators, e.g.

\[
{\cal G}_1  =  \langle ^1S|\psi^{\dag}\chi\chi^{\dag}\psi|^1S\rangle 
\]
\[
{\cal F}_1  =  \langle ^1S|\psi^{\dag}\chi\psi^{\dag} 
(\frac{-i}{2}\stackrel{\leftrightarrow}{\bf D})^2\chi |^1S\rangle
\]
\[
{\cal H}_1  = \langle ^1P|\psi^{\dag}(i/2)\stackrel{\leftrightarrow}{\bf D}
\chi.\chi^{\dag}(i/2)\stackrel{\leftrightarrow}{\bf D}\psi|^1P\rangle
\]
\[
{\cal H}_8  =  \langle ^1P|\psi^{\dag}T^a\chi\chi^{\dag}T^a\psi|^1P\rangle
\]

The subscript $8$ in ${\cal H}_8$ denotes a ``color octet'' contribution, 
which is sensitive the the $Q \overline{Q} g$ component in P-wave 
quarkonia, where the heavy quark and anti-quark form a color octet 
combination.  The remaining three singlet contributions can be 
written in terms of quarkonium wave functions at the origin or 
derivatives acting on wave functions,  using the vacuum saturation 
approximation. For instance,
${\cal G}_1/|\langle^1S_0|\psi^{\dag}\chi|0\rangle|^2  =  (1+{\cal O}(v^4))$ 
leads to $ {\cal G}_1 \sim {3 \over {2 \pi}} |R_S(0)|^2 $  ( $R$ : radial 
wave function) and 
${\cal H}_1/|\langle^1P_1|\psi^{\dag}\frac{-i}{2}\stackrel{\leftrightarrow}
                        {\bf D}\chi|0\rangle|^2  =  (1+{\cal O}(v^4)) $
 to $ {\cal H}_1 \sim { 9 \over {2 \pi}}|R_P^{\prime}(0)|^2 $ etc.  
The Argonne group finds that vacuum saturation is well satisfied. 
They work with the lowest order ( $O(M v^2)$) NRQCD action in 
their simulations.  Their results include one-loop matching between lattice 
and continuum operators.  A comparison between lattice results and 
experiment is given in Table 1.  Within large errors, one sees good 
agreement for charmonium, less so for $\Upsilon$.  Since these matrix 
elements are sensitive to wave functions at the origin, one expects 
significant quenching corrections. 
One should also note the large uncertainty coming from perturbative 
matching.  The Argonne group is looking into the possibility of 
nonperturbative renormalization of the relevant operators.

\begin{table}
\begin{tabular}{|l|c|c}
                         & lattice & experiment       \\
\hline
\hline
\multicolumn{2}{|l|}{charmonium $6/g^2=5.7$}                 &          \\ 
${\cal G}_1$                 
                              & 0.3312(6)(30)$^{+681}_{-483}$~GeV$^3$
                              & 0.36(3)~GeV$^3$                            \\
${\cal F}_1 / {\cal G}_1$  
                              & 0.07 --- 0.82~GeV$^2$
                              & 0.057~GeV$^2$                              \\
${\cal H}_1$               
                              & 0.0802(6)(77)$^{+167}_{-118}$~GeV$^5$
                              & 0.077(19)(28)                              \\
                   &          & $\;\;\;\;\;$~GeV$^5$                       \\
${\cal H}_8 / {\cal H}_1$  
                              & 0.057(1)(4)$^{+27}_{-21}$~GeV$^{-2}$
                              & 0.095(31)(34)                              \\
                   &          & $\;\;\;\;\;$~GeV$^{-2}$                    \\
\hline
\multicolumn{2}{|l|}{bottomonium $6/g^2=5.7$}                 &        \\
${\cal G}_1$                
                              & 2.129(2)(15)$^{+274}_{-218}$~GeV$^3$
                              & 3.55(8)~GeV$^3$                            \\
${\cal F}_1 / {\cal G}_1$   
                              & -6.9 --- 0.4~GeV$^2$
                              &     -----                                  \\

\hline
\multicolumn{2}{|l|}{bottomonium $6/g^2=6.0$}                  &        \\
${\cal G}_1$               
                              & 2.340(8)(19)$^{+173}_{-151}$~GeV$^3$
                              & 3.55(8)~GeV$^3$                            \\
${\cal F}_1 / {\cal G}_1$  
                              & -2.0 --- 1.6~GeV$^2$
                              &     -----                                  \\
${\cal H}_1$               
                              & 1.63(7)(23)$^{+19}_{-15}$~GeV$^5$
                              &     -----                                  \\
${\cal H}_8 / {\cal H}_1$  
                              & 0.00262(3)(24)$^{+57}_{-51}$GeV$^{-2}$
                              &     -----                                 

\end{tabular}
\caption{Continuum renormalized decay matrix elements from lattice
calculations, compared with those extracted from experimental decay rates
where available [27]. Errors in the lattice results correspond from 
left to right to statistical, systematic and perturbative matching
 errors.}
\end{table}

\section{Spin Splittings}

Spin splittings in quarkonia are sensitive to every single detail in 
the action.  They are significantly affected by quenching and have a strong 
dependence on $M_Q$.   Tadpole improvement of the action, to remove 
lattice artifacts, was found to be crucial for achieving 
 agreement with experiment. 

In Fig. 5 we show charmonium hyperfine splittings in the quenched 
approximation versus $a^2$.  There are three data points from 
Fermilab using heavy clover at $ \beta$=5.7, 5.9 and 6.1\cite{aida}. 
 The NRQCD point 
comes from simulations at $ \beta = 5.7$\cite{nrqcd2}. 
 Then there are three coarse 
lattice results using D234(3/2)\cite{klassen} on an anisotropic
 lattice plus data from 
Howard Trottier\cite{trottier}, who applied NRQCD on $O(a^2)$ improved glue.
  The ``$\times$'''s 
differ from the ``+'''s, in that the former have $a^2$ errors removed 
from the $\vec{E}$ and $\vec{B}$ fields ( this is an $a^2$ improvement 
of the NRQCD action that goes beyond tree-level, and appears to be 
noticeable for $ a > 0.2 fm$).  
It should be noted that current NRQCD simulations have errors of 
order $M_c v_c^6 \sim 35 MeV$ for $c \overline{c}$ and of order 
$M_b v_b^6 \sim 5 MeV$ for $ b \overline{b}$.  
 Very preliminary estimates of some of the 
$Mv^6$ corrections in charmonium 
indicated the hyperfine splitting could come down significantly. 
Hence, even though the current NRQCD number at $\beta = 5.7$ 
is closest to experiment, 
the true quenched result could be much lower, e.g. at about the $70 MeV$ 
level seen by the Fermilab group.  It is important that all 
the spin dependent $M v^6$ corrections be included in a future NRQCD 
simulation.

\begin{figure}
\epsfxsize=7.0cm
\centerline{\epsfbox{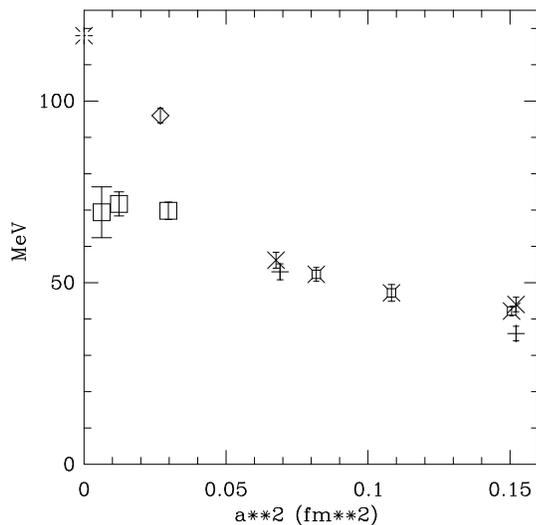}}
\caption{
{\bf Quenched Charmonium Hyperfine Splittings} versus $a^2$.  
Fermilab : boxes; NRQCD : diamond; D234(2/3)3:1 : fancy boxes; 
Trottier :$+$ \& $\times$ ;
; experiment : burst.  Errors are statistical. }

\end{figure}

\section{Summary}
Quarkonia provide many opportunities for lattice gauge theory.  There 
is a rich phenomenology of spectroscopy, decays and yet to be discovered 
states.  $\Upsilon$ S-P and 1S-2S 
splittings provide well controlled scale 
determinations that lead to accurate determinations of $\alpha_s$. Spin 
splittings still pose challenges, but we should be up to them.  
Quarkonia serve as good testing ground for new actions. 
Studies of quarkonia create invaluable experience for going 
onto heavy-light systems.

\vspace{.2in}
\noindent
\underline{Acknowledgements} \\
I thank all the people who sent results contained in this review and 
my colleagues in the NRQCD and heavy-light collaborations for 
their support and for useful conversations.  The organizers of 
LAT'96 are to be congratulated for an excellent conference.  
The author's research is supported in part by the U.S. DOE under 
DE-FG02-91ER40690.

\end{document}